\documentstyle[sprocl,epsf,epsfig]{article}

\def\beq{\begin{equation}}
\def\eeq{\end{equation}}
\def\bea{\begin{eqnarray}}
\def\beaa{\begin{eqnarray*}}
\def\eea{\end{eqnarray}}
\def\eeaa{\end{eqnarray*}}
\def\bq{\begin{quote}}
\def\eq{\end{quote}}
\def\gappeq{\mathrel{\rlap {\raise.5ex\hbox{$>$}}
{\lower.5ex\hbox{$\sim$}}}}

\def\lappeq{\mathrel{\rlap{\raise.5ex\hbox{$<$}}
{\lower.5ex\hbox{$\sim$}}}}

\def\be{\begin{equation}}
\def\ee{\end{equation}}
\def\bea{\begin{eqnarray}}
\def\eea{\end{eqnarray}}
\begin{document}

\begin{flushright}
CERN-TH/2001-086\\
ISN-01-026
\end{flushright}

\title{INVESTIGATION OF THE ROLE OF ELASTIC UNITARITY IN HIGH-ENERGY SCATTERING:  GRIBOV'S
THEOREM AND THE FROISSART BOUND}

\author{ANDRE MARTIN}

\address{TH Division, CERN, CH-1211 Geneva 23, Switzerland \\
and\\
LAPTH, BP 110, 74941 Annecy le Vieux, France\\
email: Andre.Martin@cern.ch} 
 
\author{JEAN-MARC RICHARD}

\address{Institut des Sciences Nucl\'eaires, CNRS-IN2P3, Universit\'e Joseph Fourier \\
53, Avenue des Martyrs, F-38026 Grenoble, France\\
email: jean-marc.richard@isn.in2p3.fr}


\maketitle\abstracts{ We re-examine V. Gribov's theorem of 1960 according to which the total cross-section
cannot approach a finite non-zero limit with, at the same time, a diffraction peak having a finite slope. We
are very close to proving by an explicit counter-example that elastic unitarity in the elastic region is an
essential ingredient of the proof.  By analogy, we raise the question of the saturation of the
Froissart-Martin bound, for which no examples incorporating elastic unitarity exist at the present time.}

\begin{center}
\it{Based on a talk given by one of us (A.M.) at the 
First Workshop on ``Forward Physics and Luminosity Determination at LHC", Helsinki,\\ 31 October - 4 November 2000}
\end{center}

\section{Introduction}

The work we want to present is not completely finished, but, at the time of
writing, we are 99\% convinced that the results we present are correct and will be made -- ``dans un mois,
dans un an", as Fran\c{c}oise Sagan says -- completely rigorous.

The Gribov theorem to which we refer is the statement that at high energy the scattering amplitude cannot behave
like 
\beq
s \; f(t) 
\label{1}
\eeq
for $s \to \infty$, $t$ fixed, where $s$  is the square of the centre-of-mass energy and $t$ the square of the
momentum transfer (negative or zero for $t$ physical).

In physical terms this means that the proton cannot behave like an object with a fixed size independent of the
collision energy,leading to a finite, non-zero, constant cross-section at high energy, and an asymptotically
fixed diffraction peak.  This theorem, which appeared in the proceedings of the 1960 International Conference on
High-Energy Physics \cite{aaa} (V. Gribov was not allowed to attend the conference!), came in the ``Pre-Regge"
\cite{bb} period, and was the first blow to destroy the naive beliefs of most physicists about elementary
particles at the time.  I consider it as a turning point in  ``forward physics" , the topic of this conference.

To find a contradiction, Gribov assumes not only that $F(s,t) \sim i \; s f(t)$ for $t$ physical ($i$ is
necessary for a crossing symmetric amplitude!), but also for $t$ in a complex domain.  This assumption can be weakened
a little, as we shall see later.  Also it is really done for the simple case of pion-pion scattering, but it would be
very awkward if the behaviour $F \sim s f(t)$ was forbidden for pion-pion scattering and allowed for nucleon-nucleon
scattering.

We now present Gribov's original proof, which is remarkably simple.  Here we consider the $\pi_0 \pi_0 \to
\pi_0 \pi_0$ amplitude where the pion has isospin zero.  This is an \underline{inessential} simplification.
To generalize to the real pion, which is an isospin triplet, one can consider simultaneously all possible
channels or play with the fact that, strictly speaking, $m_{\pi_0} < m_{\pi^\pm}$.  In Fig. 1 we have
represented the Mandelstam diagram with
$m_{\pi} = 1$. 
\begin{figure}
\begin{center} 
%\vspace*{2cm}
\epsfig{figure=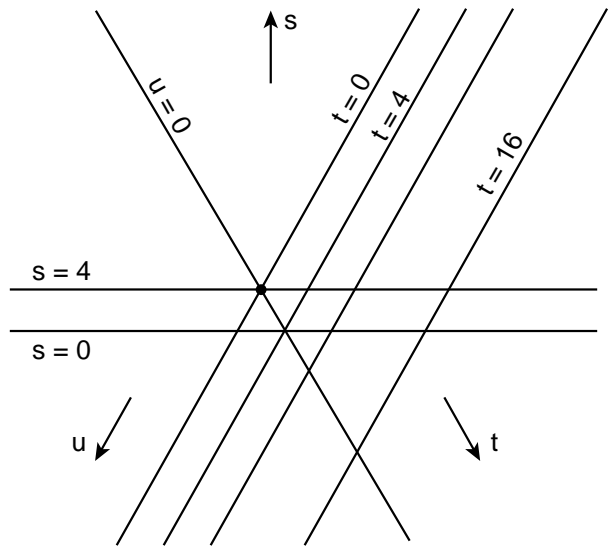,width=8cm}
\caption{Mandelstam diagram}
 \end{center}
\end{figure}
We assume Mandelstam representation where the scattering amplitude is 
\bea
\frac{1}{\pi^2} \, \int \int \, \frac{
ds' dt' \rho(s', t')}
{(s' - s)(t' - t)} +
{\rm circular \, permutations \, on} \, s, t, u \\
+ {\rm subtractions} \nonumber
\label{2}
\eea
with $ s + t + u = 4$.

In fact we use a much smaller domain of analyticity than Mandelstam (for a method of construction of
such domains, see for instance Ref. \cite{cc}).  However, the domain obtained from axiomatic field
theory and unitarity \cite{cc} is not sufficient.

We assume $F(s,t) \sim sf(t)$, but from crossing in the $s \rightleftharpoons u$ variable we must have
\beq
F(s,t) \sim i \; s \, f(t),
\label{3}
\eeq
$f(t)$ real for $t \leq 0$, so that the asymptotic absorptive part is just $s f(t)$, where $f(t)$ must
have a cut from $t = 4$ to $\infty$, and the double spectral function is given by
\beq
\rho (s,t) \simeq s \, Im \, f(t).
\label{4}
\eeq

Now Gribov uses elastic unitarity in the elastic strip of the $t$ channel which, originally, reads:
\beq
A_t (t, z_t) = 
\int K (t, z_t, z'_t, z''_t)
F(t, z'_t) \, F(t, z''_t) \;
dz'_t dz''_t
\label{5}
\eeq
where $z_t = 1 + 2s/(t-4)$ is the cosine of the scattering angle in the $t$ channel.  Equation (5) is
just a fancy way to write a spherical convolution of $F$ and $F^*$. $K$ is a known kernel, the square
root of a polynomial whose support is entirely in the physical region.

Because of Mandelstam representation, $A_t(t, z_t), F(t, z_t), F^*(t, z^*_t)$ are all analytic in a
cut plane in $z_t$, with cuts at $z_t = \pm (1 + 8/(t-4))$, and Equation (5), as shown by Mandelstam,
can be continued on the cuts, and transformed into  a relation connecting the discontinuities of $A_t, F$
and $F^*$:
\beq
\rho (s,t) = 
\int \hat{K} (t,s,s_1, s_2)
A_s (s_1, t) A^*_s (s_2, t) ds_1 ds_2
\label{6}
\eeq

$\hat{K}$ is again a completely known expression with a finite support in $s_1, s_2$ for given $s$ and
$t$.

When $s \to \infty$, the domain of integration grows in such a way that the maximum of $s_1$ and $s_2$
also tends to infinity, and it is not difficult to convince oneself that if $A_s \sim s \, f(t)$, the
dominating region of the integral is such that both $s_1$ and $s_2 \to \infty$.  It is not difficult
to see that the right-hand side of Eq. (6) behaves like
\beq
|f(t)|^2 \, s \, \log \,s
\label{7}
\eeq
while the left-hand side is obviously
\beq
\rho = s \, Im \, f(t)
\label{8}
\eeq
This is the Gribov contradiction, which occurs also for $F \sim s/ \log \, s$.  Only if $F \sim
s/(log \, s)^{1 + \epsilon}$ is the contradiction removed.  This was the way out that Gribov proposed,
corresponding to a total cross-section going to zero, but Nature did not make this choice.  As we have
known since 1972, cross-sections are rising \cite{dd} and might go to infinity at infinite energy.

Another way of looking at the Gribov phenomena is to use the complex angular momentum plane.  In this
language, the Gribov theorem says that elastic unitarity in the $t$ channel forbids a ``fixed pole" at
$J = 1$.  It is slightly more general in the sense that it is clearly \underline{insufficient} to take
an amplitude
\beq
i \, s \, f(t) + g(t) s^{\alpha (t)}
\label{9}
\eeq
with $\alpha (t) < 1$ for $t \leq 0$ and $ Re \alpha (t) > 1$ for $t > 4 m^2_{\pi}$ to ``turn" the
Gribov Theorem.  In this version one clearly sees that the Donnachie-Landshoff odderon\cite{ee}, $s
\, t^{-4}$, is also, strictly speaking, forbidden unless there is an insurmountable natural boundary in the
complex angular momentum plane.

One of the reasons for our renewed interest in the Gribov theorem is that there is another problem
concerning high-energy cross-sections.  As we said, we know that they are rising, possibly to
infinity, possibly saturating \underline{qualitatively}\cite{dd} the Froissart-Martin bound\cite{ff}
\beq
\sigma_t < C(\log \, s)^2
\label{10}
\eeq

The question is to know if this bound can be saturated if one takes into account unitarity
constraints. The simplest unitarity constraint is:
\beq
Im \, f_{\ell} (s) \geq 
\frac{2k}{\sqrt{s}} \, \left |f_{\ell}(s) \right |^2
\label{11}
\eeq
when $f_{\ell}(s)$ is a partial wave amplitude.  It is valid at \underline{all} energies.

Joachim Kupsch has been able to show the existence of amplitudes which satisfy Mandelstam
representation and the general unitarity constraint (11), and saturate the bound (10) \cite{ggg}. This
was a formidable achievement, taking into account all the restrictions that amplitudes saturating
(10) must satisfy, in particular the so-called AKM conditions \cite{hh}. 

However, it is legitimate to ask oneself if the elastic unitarity condition in the elastic region 
\beq
Im \, f_{\ell}(s) = 
\frac{2k}{\sqrt{s}} \, \left |k_{\ell} (s) \right |^2,
\label{12}
\eeq
which is so constraining in the Gribov case, might prevent the qualitative saturation of the Froissart
bound.  Turning the argument upside down, we may ask if the Gribov theorem survives if we require only
(11) but not (12).

We want to show, and have almost succeeded in doing so, that there are explicit examples of amplitudes
behaving like $s \, f(t)$ i.e., violating Gribov's theorem, satisfying the general condition (11) at
all energies.  If this is so, all doubts are allowed concerning the other problem, the saturation of
the Froissart bound because, while D. Atkinson succeeded in	1970 \cite{jj} to construct amplitudes
satisfying (11) and (12) but with cross-sections decreasing faster than $(\log \,s)^{-2}$, nobody has
succeeded in saturating the Froissart bound at the same time.

If it were to be shown one day that (12) prevents the qualitative saturation of the Froissart bound,
it would be a major blow to many models such as the one of Bourrely-Cheng-Soffer-Walker-Wu\cite{kk}, where 
$\sigma_t
\sim (\log \, s)^2$ at extreme energies.  I admit that the likelihood of this happening is very small,
first of all because it requires somebody with great classical mathematical skill, who would tend
naturally to use his abilities on a more fashionable subject.  Nevertheless, we are allowed to raise
the question.

\section{A Candidate for Violating Gribov's Theorem}

We propose to take
\beq
F = C [F_s + F_t + F_u ]
\label{13}
\eeq
\bea
F_s & = & (4 - \sqrt{4-t}\sqrt{4-u}) \exp -2(4-s)^{1/4}\nonumber \\
F_t & = & (4 - \sqrt{4-u}\sqrt{4-s}) \exp -2(4-t)^{1/4}\nonumber \\
F_u & = & (4 - \sqrt{4-t}\sqrt{4-s}) \exp -2(4-u)^{1/4}
\label{14}
\eea
For $s \to \infty$, $t$ fixed, $F_t$ dominates and behaves like $i \, s \, \exp -2(4-t)^{1/4}$, which
is what we want.  We must now show that (11) is satisifed and, first of all, that all partial waves
have a \underline{positive} imaginary part. It is possible to show that in this channel $Im(F_t +
F_u)$ has positive partial wave amplitudes, because $Im(F_t + F_u)$ can be expanded in powers of $\cos
\theta_s$:
\beq
Im(F_t + F_u) = \Sigma \, C_n(\cos \theta_s)^n,
\label{15}
\eeq
and it can be shown that the $C_n$'s are \underline{positive}.

Then, since
\beq
(\cos \theta)^n = \Sigma \, C_{\ell, n} \, P_{\ell} (\cos \theta)
\label{16}
\eeq
with $C_{\ell, n} > 0$, positivity of the partial waves follows.  Another consequence from (15) and
(16) is that the partial waves of $Im (F_t + F_u)$ are \underline{decreasing}.  However, $ImF_s$ has
oscillating signs, and though, globally, it is small in the $s$ channel, it must be proved that it is small
partial wave by partial wave.

There are limiting cases where the behaviour is correct:\\
\\
\noindent
 i) $\ell$ fixed, $k \to \infty$ and, in fact, also $k \to \infty$ and $\ell \to \infty$ with $\ell /
k \to 0$, which allows $\ell \to \infty$ in some way.  Then from the direct integration of $F$ with
$P_{\ell} (\cos \theta_s)$ one gets:
\bea
f_{\ell} & \simeq & \left ( \frac{8}{s} + 2i \right ) \, \int^{0}_{- \infty} \,
dt \, \exp \, -2(4-t)^{1/4} \nonumber \\
& \simeq & \left ( \frac{4}{s} + i \right ) \, (15 + 14 \sqrt{s} ) \, \exp \, -3 \sqrt{2}
\label{17}
\eea
This shows that ``unitarity" is satisfied just by adjusting $C$ in (13).\\
\\
\noindent
ii) $k$ fixed, $\ell \to \infty$ and more generally $\ell / k \to \infty$, $\ell \to \infty$.  Then
one has to use the Froissart-Gribov representation of the different contributions of the partial waves
of the form
\beq
x_{\ell} = \int^{\infty}_{u} \, w(s,t) \, Q_{\ell} \left ( 1 + \frac{t}{2k^2} \right ) \,
\frac{dt}{2k^2}
\label{18}
\eeq
If
\beq
w \sim C(s) (t-4)^{\nu}
\label{19}
\eeq
we get for $\ell \to \infty$, $\ell /k \to \infty$:
\beq
x_{\ell} \sim C(s) \left ( \frac{2k\sqrt{2}}{\ell + 1} \right )^{\nu + 1} \,
\Gamma (\nu + 1) \,
Q_{\ell} \left ( 1 + \frac{2}{k^2} \right )
\label{20}
\eeq
In this way it is trivial to see that the contribution from $Im (F_u + F_t)$ dominates all the rest
and that unitarity and positivity are satisfied.\\
\\
\noindent
iii) Finally, in the ``scaling limit"
\beq
\frac{\ell + 1}{k} = b, \, b \, {\rm fixed} \, k \to \infty
\label{21}
\eeq
one can use the asymptotic expression for the $Q_{\ell}$'s valid for large $\ell$, uniform in $\phi$:
\beq
Q_{\ell} (\cosh \phi ) \simeq \sqrt{\frac{\phi}{\sinh \phi}} \,K_0 \left( (\ell + 1/2) \phi \right)
\label{22}
\eeq
The RHS contribution to $Im F^t_{\ell}$ becomes:
\beq
\frac{1}{2k^2} \, \int \, dt \, K_0(b \sqrt{t}) \,
\sin \sqrt{2} (t-4)^{1/4} \, \exp - \sqrt{2} (t-4)^{1/4},
\label{23}
\eeq
and is necessarily positive, because of the positivity properties of $Im (F_t + F_u)$, and necessarily
dominates all the other terms, the real parts and the contribution from $Im F_s$ which decrease exponentially
with $s$.\\
\\
\noindent
iv) Unitarity of the $\ell = 0$ partial wave can be checked by hand. There, in the expression $(4 -
\sqrt{4-t}\sqrt{4-u})$, the cancellation at $s = 4$ is essential.

We conclude that it is only necessary to check unitarity in a \underline{finite} region of the $\ell, k$
domain.  First we did this \underline{numerically}.  Tests, at selected energies where unitarity seems
endangered by the negative signs of the partial wave expansion of $Im F_s$, indicate that provided one
divides the amplitude by a factor 2.1, there does not seem to be any violation.  These tests are
presented in Table 1.

However, we believe that by using old and new properties of the $Q_{\ell}$'s, for instance 
\beq
{\rm - old:}\, Q_L(z) / Q_{\ell} (z) , {\rm decreases \, with} \; z \;{\rm for} \, L > \ell
\label{24}
\eeq
\beq
{\rm - new:} \, Q_{M-1} \left ( 1 + \frac{t}{2k^2} \right ) / Q_{m-1} \,
\left ( 1+ ( \frac{M}{m} )^2 \, \frac{t}{2k^2} \right ) \,
{\rm decreases \, with} \, t \,{\rm for} \, M > m
\label{25}
\eeq
We can complete the proof analytically, except for the calculation of a \underline{small}, finite
number of integrals over a \underline{finite} interval.

Let us illustrate by two examples how we use (24) and (25).  One problem is to find a lower bound on 
\begin{eqnarray*}
&&Im F^{\ell}_{t, RHC} =\\
&& \frac{1}{2k^2}
\int\limits^{\infty}_{4} \, Q_{\ell} \left ( 1 + \frac{t}{2k^2} \right ) \,
\sqrt{s-4} \sqrt{s+t} \,
\sin \sqrt{2} (t-4)^{1/4} \, \exp - \sqrt{2} (t-4)^{1/4} \, dt.
\end{eqnarray*}
With the change of variable $x = \sqrt{2} (t-4)^{1/4}$, the integral becomes
$$
\int^{\infty}_{0} \, W(x,s) \, sin \, x \, dx.
$$
It is easy to prove that for $x > 2 \pi$, $W(x,s)$ is decreasing in $x$.  Hence
$$
\int^{\infty}_{2 \pi} \, W(x,s) \, \sin \, x \, dx > 0.
$$
So
\begin{eqnarray*}
&&Im F^{\ell}_{t,RHC} > g_{\ell}(s) = \\
&&\frac{1}{2k^2}
\int\limits_{4}^{4(1+ \pi^4)} \!\!\!\!\!\! Q_{\ell} \left (1 + \frac{t}{2k^2} \right ) \,
\sqrt{s-4} \sqrt{s+t} \, \sin \sqrt{2} (t-4)^{1/4} \, 
\exp - \sqrt{2} (t-4)^{1/4} \, dt.
\end{eqnarray*}
In $g_{\ell}$ the integrant has a single change of sign at $t_0 = 4 + \frac{\pi^4}{4}$. If
$g_{\ell_0}(s_0) > 0$, it follows from (24) that
$$
g_{\ell}(s_0) > g_{\ell_0}(s_0) \,
\frac{Q_{\ell_0} \left (1 + \frac{t_0}{2k^2} \right )}{Q_{\ell} \left( 1 + \frac{t}{2k^2} \right)}
$$
if $\ell > \ell_0$, and hence we get a lower bound on $Im f^{\ell}_{t,RHC}(s_0)$ for $\ell > \ell_0$,
but we can also change at the same time $\ell$ and $s$ and use (25) to get a lower bound for 
$$
\frac{\ell + 1}{2k^2} = \frac{\ell_0 +1}{2k^2_0}, \, \ell > \ell_0.
$$

Finally, we can also use the fact that $Im \left( f^{\ell}_{t} + f^{\ell}_{u} \right )$ is decreasing
with $\ell$ to get lower bounds for smaller $\ell$'s.

Everybody will understand that this is a rather delicate book-keeping, but with enough patience, we
will succeed.

\section{Conclusions}

If you believe that our example, after ``renormalization",
satisfies the general unitarity condition (11), you conclude that the proof of Gribov's theorem holds
only if we impose elastic unitarity (12) in the elastic region.  Incidentally, a kind of universality
postulate is tacitly made by everybody, because  everybody believes that the Gribov theorem applies to
$pp$ scattering, while technically there are problems with the unphysical region in the $t$ channel which
would disappear only if $m_p < 1.5 m_{\pi}$.

It is therefore legitimate to ask oneself what happens in another situation, the qualitative
saturation of the Froissart bound, where the examples saturating this bound only satisfy condition
(11) but not elastic unitarity.  Would elastic unitarity make the saturation of the bound impossible?

My own prejudice is that this will not happen, but there is absolutely no proof of that.  As pointed
out by S.M. Roy, \cite{lll} it could be that elastic unitarity leads to a \underline{quantitative}
improvement of the Froissart bound, i.e., replacing the factor $\pi / m^2_{\pi}$, which appears in the
Lukaszuk-Martin \cite{mm} version:
$$
\sigma_t < \frac{\pi}{m_{\pi^2}} \, (\log s)^2,
$$
by something much smaller.  In any case, we believe that these questions are worth investigating.

\section{Acknowledgements}

One of us (A.M.) would like to thank J. Kupsch and S.M. Roy for very stimulating discussions, and the Indo-French
Centre for the Promotion of Advanced Research (IFCPAR) for support.

\begin{table}[t]
\caption{Numerical tests of unitarity of (13), with $C = 1$}
\vspace{0.2cm}
\begin{center}
\footnotesize
\begin{tabular}{|c|c|c|c|c|}
\hline
$s$ & $\ell$ & $Re f_{\ell}$ & $Im f_{\ell}$ & $\sqrt{s}/2k \; Im f_{\ell}/|f_{\ell}|^2$\\
\hline
4.0001 & 0 & 0.473 & 0.00236 & 2.1\\
&2 & 4.1 $\times  10^{-11}$ & 5.1 $\times 10^{-12}$ & 6 $\times 10^{11}$\\
\hline
4.01 & 0 & 0.108 & 0.0040 & 2.19 \\
& 2 & 7 $\times 10^{-6}$ & 1.19 $\times 10^{-7}$ & 4 $\times 10^{5}$ \\
& 4 & 8.5 $\times 10^{-11}$ & 9.6 $\times 10^{-11}$ & 3.7 $\times 10^{10}$\\
\hline
50 & 0 & 0.53 & 1.45 & 0.63 \\
& 2 & 0.037 & 1.19 & 0.84 \\
& 4 & 0.0081 & 0.64 & 1.56 \\
\hline
70 & 0 & 0.40 & 1.71 & 0.57 \\
Chosen & 2 & -0.009 & 0.294 & 3.5 \\
so that & 4 & 0.002 & 0.0736 & 14 \\
$Im f^{s}_{\ell} < 0$ & 6 & 0.0008 & 0.0193 & 53 \\
for $\ell > 0$ & 8 & 0.00025 & 0.0054 & 190\\
\hline
3500 & 0 & 0.0055 & 2.06 & 0.49 \\
& 2 & 0.0017 & 1.82 & 0.54 \\
& 4 & 0.0017 & 1.51 & 0.66\\
idem & 6 & 0.0014 & 1.21 & 0.82\\
& 8 & 0.0011 & 0.96 & 1.04 \\
& 10 & 0.0009 & 0.76 & 1.32 \\
& 12 & 0.0007 & 0.60 & 1.67 \\
& 14 & 0.0005 & 0.477 & 2.1 \\
& 16 & 0.0004 & 0.380 & 2.62 \\
& 18 & 0.0003 & 0.305 & 3.3 \\
& 20 & 0.0003 & 0.245 & 4.1 \\
\hline
24000 & 0 & 0.0003 & 2.05 & 0.49 \\
& 2 & 0.0003 & 2.01 & 0.49 \\
& 4 & 0.0003 & 1.94 & 0.51 \\
& 6 & 0.0003 & 1.84 & 0.54 \\
& 8 & 0.0003 & 1.72 & 0.58 \\
& 10 & 0.0003 & 1.60 & 0.63 \\
& 12 & 0.0002 & 1.47 & 0.68 \\
& 14 & 0.0002 & 1.35 & 0.74 \\
Idem & 16 & 0.0002 & 1.24 & 0.80 \\
& 18 &0.0002 & 1.13 & 0.88 \\
& 20 & 0.0002 & 1.04 & 0.96 \\
& 22 & 0.0002 & 0.95 & 1.05 \\
& 24 & 0.0001 & 0.81 & 1.15 \\
& 26 & 0.0001 & 0.79 & 1.26 \\
& 28 & 0.0001 & 0.72 & 1.37 \\
& 30 & 0.0001 & 0.66 & 1.50 \\
& 32 & 0.0001 & 0.61 & 1.65 \\
& 34 & 0.0001 & 0.56 & 1.80 \\
& 36 & 0.0001 & 0.51 & 1.97 \\
& 38 & 0.0001 & 0.47 & 2.15 \\
& 40 & 0.0001 & 0.43 & 2.34 \\
\hline
$\infty$ & Any finite & 0 & (15 + 14 $\sqrt{2}) \times$ & 0.486187 \\
& $\ell$ & & $\exp - 2 \sqrt{2}$ & \\
& & & $\simeq 2.05602$ & \\
\hline

\end{tabular}
\end{center}
\end{table}


\begin{thebibliography}{99}
\bibitem{bb}
The papers by Regge on Regge poles in potential scattering did exist, but the application
to high-energy scattering was not yet fashionable.
\end{thebibliography}
\end{document}